\documentstyle[12pt,fleqn,epsfig]{article} 
\def\be{\begin{eqnarray}}
 \def\ee{\end{eqnarray}} 
\begin{document} 
 \pagestyle{empty}
 \Huge{\noindent{Istituto\\Nazionale\\Fisica\\Nucleare}}

\vspace{-3.9cm}

\Large{\rightline{Sezione SANIT\`{A}}} \normalsize{} \rightline{Istituto
Superiore di Sanit\`{a}} \rightline{Viale Regina Elena 299} \rightline{I-00161
Roma, Italy}

\vspace{0.65cm}

\rightline{INFN-ISS 97/12} \rightline{October 1997}

\vspace{0.5cm} 
\renewcommand{\textfraction} {.01} 
\renewcommand{\topfraction} {.99}

\font\larger=cmr12 scaled\magstephalf 
\font\Larger=cmr14 scaled\magstephalf

\vspace{2cm}

\begin{center} { \Large \bf {Can neutron electromagnetic form factors be
obtained by polarized inclusive electron scattering off polarized three-nucleon
bound states?}}

\vspace{1cm} 
{\larger{ A.  Kievsky$^a$, E.  Pace$^b$, G.  Salm\`e$^c$ and M.
Viviani$^a$}}\\ 
\vspace{0.5cm} 
{\small{\it{$^a$INFN, Sezione di Pisa, 56010
S.Piero a Grado, Pisa, Italy } \\ 
{$^b$Dipartimento di Fisica, Universit\`{a} di
Roma "Tor Vergata" and Istituto Nazionale di Fisica Nucleare, Sezione Tor
Vergata, Via della Ricerca Scientifica 1, I-00133 Roma, Italy}
\\ {$^c$INFN,
Sezione Sanit\`{a}, Viale Regina Elena 299, I-00161 Roma, Italy}}}

\end{center} 
\vspace{2cm} 
\begin{abstract} The investigation of the
electromagnetic inclusive responses of polarized $^{3}$He within the plane wave
impulse approximation is briefly reported.  A particular emphasys is put on the
extraction, from the inclusive responses at the quasielastic peak, of the
neutron form factors from feasible experiments.  
\end{abstract} 

\vspace{3cm}
\hrule width5cm 

\vspace{.2cm} 

\noindent{\footnotesize{ Presented to XVth
International Conference on "Few-body problems in Physics", Groningen, July
1997.  To appear in Nucl.  Phys.  A} 
\newpage 
\pagestyle{plain}
\larger 
\section{INTRODUCTION}

\indent In the past few years the possibility of extracting the neutron
electromagnetic (em) form factors from measurements of inclusive scattering of
polarized electrons by polarized $^{3}$He has attracted great attention
\cite{Jon}.  Unfortunately, a reliable determination of neutron form factors is
prevented by many difficulties, primarily by the presence of the proton and the
effect of the final state interaction (FSI) and possibly by meson exchange
currents, isobar components in the target wave function and relativistic
effects.  In particular the overwhelming role played by the proton in the
transverse-longitudinal response has not allowed an accurate estimate of the
neutron electric form factor at low momentum transfer.

Our investigation \cite{KPSV} has been focused on the possible extraction of
both the electric and magnetic neutron form factors from the inclusive polarized
cross sections within the Plane Wave Impulse Approximation (PWIA).  We
restricted the analysis at the quasielastic peak, where FSI should play a minor
role (cf.  the preliminary calculations including FSI shown in \cite{Golak}) and
we considered a range of momentum transfer up to 2$~(GeV/c)^2$, namely the
kinematical region relevant for TJNAF \cite{TJNAF}.  \section{RESULTS} The main
ingredient of our calculations is a refined spin-dependent spectral function
\cite{KPSV} of the three-nucleon system.  Various spectral functions
corresponding to different, realistic two- and three-body nuclear forces
\cite{Av14,TM} have been considered.  As well known \cite{CPS,Sauer} the
spin-dependent spectral function yields the probability distribution of finding
a nucleon, inside a nucleus, with given polarization, three-momentum and removal
energy.  First of all, we have analyzed in detail, at the qe peak, the model
dependence upon the initial state interaction of the inclusive responses as a
function of $Q^2$, obtaining a very encouraging result (from the point of view
of a model independent determination of the neutron form factors):  the changes
due to a different choice of two- and three-body forces and to the Coulomb
interaction as well, are of the order of a few percent.  In Fig.  1, the
calculations illustrating this result for the transverse, $R^{^3He}_{T'}$, and
the transverse-longitudinal, $R^{^3He}_{TL'}$ responses at the qe peak are shown
(the Galster \cite{Gal} nucleon form factors have been adopted).  It is worth
noting that the Brazil \cite{BR} and the Tucson-Melbourne \cite{TM} three-body
interactions essentially yield the same results.  \begin{figure}
\epsfig{file=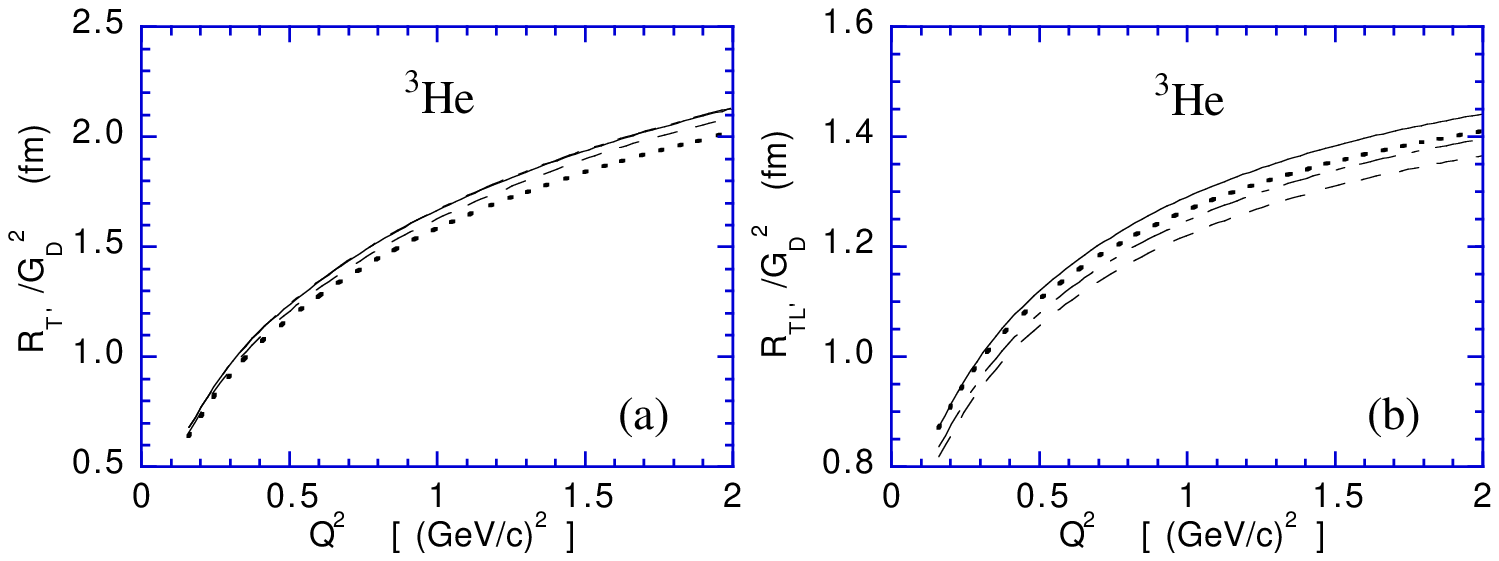,bbllx=10mm,bblly=220mm,bburx=130mm,bbury=284mm}

\parbox{16cm}{Figure 1.  a) The ratio $R^{^3He}_{T'}\left(Q^2,\nu_{peak}
\right)/G_D^2(Q^2)$ vs $Q^2$ ($G_D(Q^2)= 1/(1+Q^2/0.71)^2$).  Solid line:
$R_{T'}$ obtained from AV14 \cite{Av14} + Coulomb interaction; dot-dashed line:
the same as the solid line, but without Coulomb interaction, (this line largely
overlaps with the solid one); dotted line:  the same as the solid line, but for
RSCv8 potential \cite{Reid}; dashed line:  the three-body interaction
\cite{BR,TM} has been added to Av14 + Coulomb interaction (see text).  b) The
same as Fig.  1(a), but for $R^{^3He}_{TL'}\left(Q^2,\nu_{peak} \right)$.
(After Ref.  2)} \end{figure}

After investigating the effects of changing the interaction in the initial state
we have analyzed the proton contribution.  It should be pointed out that the
presence of the proton has a quite different effect in the inclusive responses
of polarized $^3$He:  in particular in the transverse response it is
$\approx$10$\%$, while in the transverse-longitudinal one it ranges between
80$\%$, at low values of $Q^2$, and 40$\%$, at the highest values.  These
features are illustrated in Fig.  2a.  In Fig.  2b, for the sake of comparison,
the same analysis is shown for the polarized $^3$H, that represent a very good
effective proton target, while the corresponding statement "the polarized $^3$He
is an effective neutron target" should be considered with some care for the case
of $R^{^3He}_{TL'}$.  Nevertheless the proton predominance can yield an
unexpected help if the PWIA represents a reliable approximation for
$R^{^3He}_{TL'}$ at the qe peak (cf.  \cite{Golak}).  As a matter of fact, as it
was shown in \cite{KPSV} $\partial [R^{^3He}_{TL'}/G_{M}^p G_{E}^p] /\partial
Q^2$ is almost constant at low momentum transfer (0.1 $\le Q^2 \le $0.3
$(GeV/c)^2$), and it is proportional to a neutron structure function.  This
latter quantity contains only the neutron spin-dependent spectral function,
without any em form factor dependence (the same feature occurs for the
corresponding proton structure function).  In Fig.  3, the ratio of the
polarized transverse-longitudinal response and the proton form factors, i.e.
$R^{^3He}_{TL'}\left(Q^2,\nu_{peak} \right)/\left(- 2\sqrt{2}~G^{p}_M\left(Q^2
\right)G^{p}_E\left(Q^2 \right)\right)$, is presented.  It should be pointed
that a constant behaviour of the proton contribution to such a ratio is
predicted by the PWIA, while the linear behaviour in $Q^2$ for the neutron one
is the combined effect of the almost constant value of the neutron structure
function (as shown by PWIA calculations) and the general property of the
electric neutron form factor to be linearly vanishing for $Q^2~\rightarrow~0$.
The comparison with the experimental data can allow one to extract the proton
contribution to $R^{^3He}_{TL'}$ and then to isolate the neutron one.
Furthermore the comparison could give valuable information on the extent to
which i) PWIA holds and ii) the extraction of the neutron form factors can be
achieved (or in other words, the factorization of the nucleon structure
functions can be applied \cite{KPSV}).

One can also determine experimentally the proton contribution to
$R^{^3He}_{T'}$.  In fact, a measurement of the ratio
$R^{p}_{T'}\left(Q^2,\nu_{peak} \right)/$ $R^{p}_{TL'}\left(Q^2,\nu_{peak}
\right)$ can be obtained through an accurate determination of the polarization
angle $\beta_{critic}$, where the proton contribution to the polarized cross
section, is vanishing.  An estimate of such an angle, at the qe peak, can be
obtained by measuring the integrated amount of protons knocked out along the
direction of $\vec{q}$, since at the qe peak protons are emitted preferably
along such a direction.  In principle this is an exclusive measurement, but it
can be performed more easily than the one where knocked out neutrons are
measured since i) only protons have to be detected and ii) row data are
sufficient for estimating $\beta_{critic}$ (one has only to determine the
polarization angle where the polarized contribution to the inclusive cross
section changes sign).  Finally, the PWIA prediction of this angle could be used
as a reliable guideline for the experimental measurements, since
$\beta_{critic}$ depends upon the ratio of responses, and therefore it should be
less sensitive to various effects, such as FSI, than each response separately.

\begin{figure}[t] \epsfig{figure=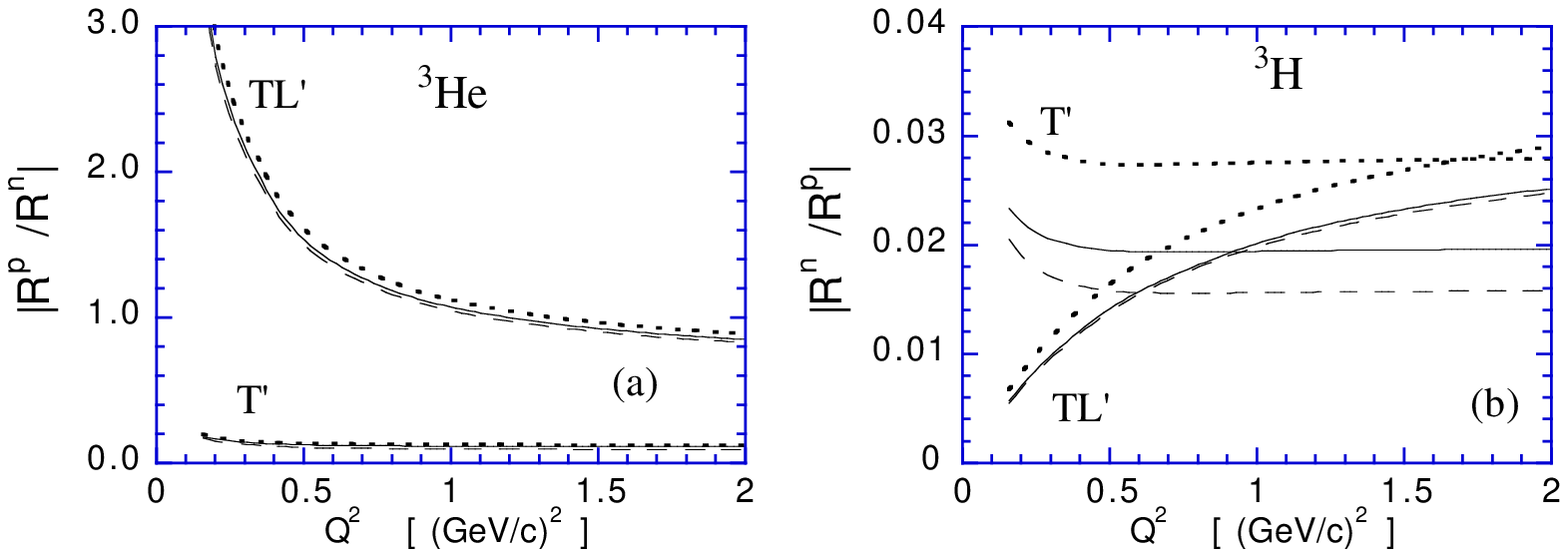,bbllx=5mm,bblly=210mm,bbury=285mm}
 \parbox{16cm}{\larger{Figure 2.  a) The ratio $R_p/R_n$ in $^3$He vs $Q^2$, for
 the transverse-longitudinal ($TL'$) and the transverse ($T'$) responses.  Solid
 lines:  inclusive responses obtained from AV14 + Coulomb interaction; dotted
 lines:  the same as the solid lines, but for RSCv8 potential; dashed lines:
 the three-body interaction has been added to Av14 + Coulomb interaction.  b)
 The same as Fig.  2(a), but for $R_n/R_p$ in $^3$H}} \end{figure}

\begin{figure}[t] \epsfig{figure=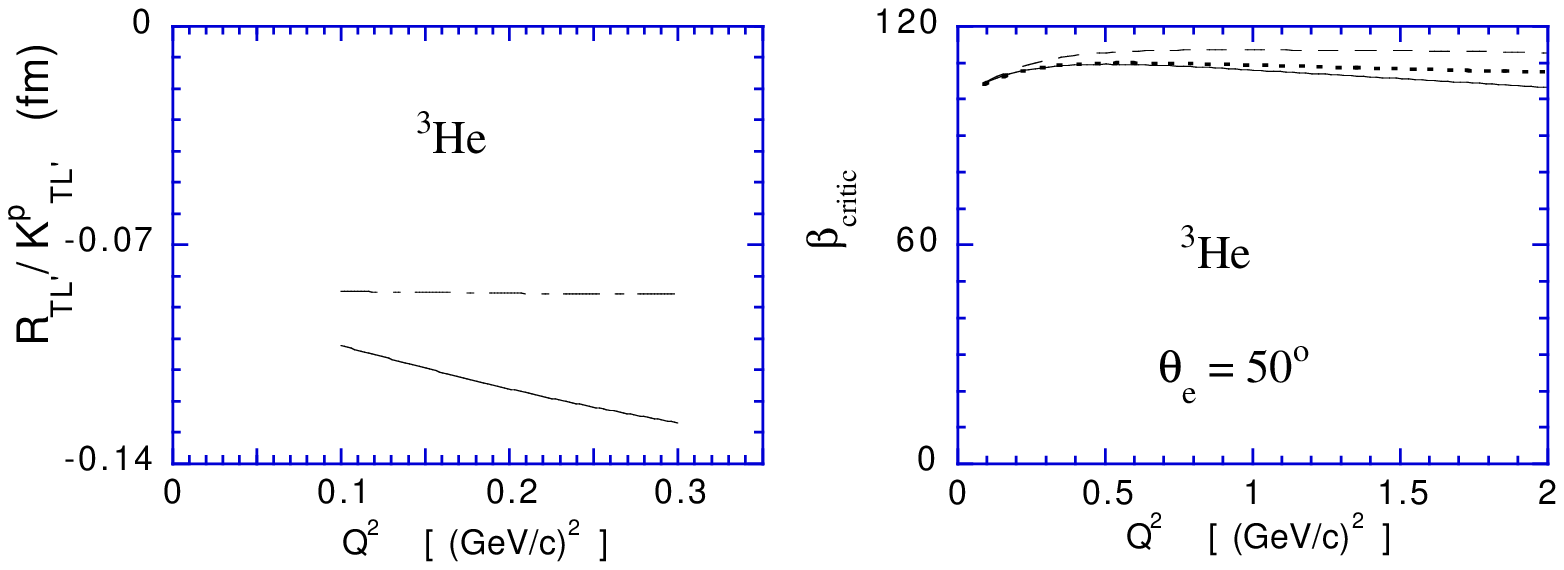,bbllx=5mm,bblly=205mm,bbury=285mm}
 \parbox{7.9cm}{\larger{Figure 3.  The ratio $R^{^3He}_{TL'}\left(Q^2,\nu_{peak}
 \right)/K^{p}_{TL'}$, with $ K^{p}_{TL'}=- 2\sqrt{2}G^{p}_EG^{p}_M$, vs $Q^2$.
 Solid line:  result obtained by the spin-dependent spectral function
 corresponding to Av14 + Coulomb interaction and the Galster nucleon form
 factors \cite{Gal}.  Dot-dashed line:  $R^{^3He,p}_{TL'}/K^{p}_{TL'}$, shown as
 a reference line.  (After \cite{KPSV}).}} $~$ \parbox{7.5cm}{\larger{Figure 4.
 The polarization angle $\beta_{critic}$, where the proton contribution to the
 polarized cross section of $^3$He, at the qe peak, vanishes, vs $Q^2$.  Solid
 line:  Av14 + Coulomb interaction; dashed line:  Av14 + Coulomb interaction +
 three-body forces; dotted line:  RSCv8 + Coulomb interaction.  (After
 \cite{KPSV}).}} \end{figure}

\section{SUMMARY} We have proposed distinct measurements that allow the
extraction of the neutron contribution to the total responses and then an
estimate of the ratio $G_{E}^{n}/ G_{M}^{n}$.  Moreover, by introducing
theoretical calculations of the nuclear structure functions one could even
obtain $G_{E}^{n}$ and $G_{M}^{n}$ separately.  It should be emphasized that a
measurement of the em inclusive responses of polarized $^3$H could allow one to
check more directly the reaction mechanism, in particular the factorization of
the responses at the qe peak (essential for extracting the neutron form
factors), and to obtain the nuclear structure functions.  Once such quantities
were available, the model dependence in the extraction of $G_{E}^n$ and
$G_{M}^n$ could be substantially lowered by using the $^3$H structure functions
as an experimental estimate of the corresponding quantities for $^3$He.

\end{document}